\documentclass[twocolumn]{article}
\usepackage{color,epsfig,amsmath,amssymb,graphicx}

\def\be{\begin{equation}}
\def\fe{\end{equation}}
\def\rf{(\ref }
\def\fr{)\,}
\def\citep{\cite}

\definecolor{red}{rgb}{0.8,0,0}
\definecolor{violet}{rgb}{0.4,0,0.4}
\definecolor{vert}{rgb}{0,0.5,0.0}
\definecolor{navy}{rgb}{0.0,0.0,0.6}
\definecolor{orange}{rgb}{0.8,0.2,0.0}
\definecolor{bleu}{rgb}{0.3,0.0,0.8}

\def\rmn{ {\rm n}} \def\rmp{{\color{orange}\rm p}}
\def\ZZ{ {\color{orange} Z}}
\def\rmf{ {\rm f}} \def\rmb{ {\rm b}} \def\rmc{ {\rm c}}
\def\rmI{ {\rm I}} \def\rmfc{ {\rm fc}}

\def\nn{{\color{vert}n}} \def\nc{\nn_{\rmc}} 
\def\nf{\nn_{\rmf}}
\def\nI{\nn_{\rmI}}

\def\calA{{\color{red}{\cal A}}}
\def\cC{{\color{red}{C}}}
\def\calE{{\color{red}{\cal E}}}
\def\rrho{{\color{red}\rho}}
\def\TT{{\color{red}T}}\def\UU{{\color{red}U}}
\def\SeS{{\color{red}S}}
\def\mm{{\color{red}m}} 

\def\ff{{\color{red}f}}
\def\hhbar{{\color{red}\hbar}}

\def\calF{{\color{red}{\cal F}}}
\def\calC{{\color{red}{\cal C}}}

\def\period{{\color{violet} P}}
\def\Om{{\color{violet}\Omega}}
\def\Oc{\Om_{\rmc}}
\def\Of{\Om_{\rmf}}
\def\ww{{\color{red} w}}

\def\elh{{\color{red}{\ell}}}
\def\hhbar{{\color{red}{\hbar}}}

\def\uu{{\color{violet}u}} \def\vv{{\color{violet}v}}
 
\def\calK{{\color{violet}{\cal K}}}

\def\hh{{\color{bleu}\varrho}}
\def\ttheta{{\color{bleu}\theta}}
\def\rr{{\color{bleu}r}} \def\nnu{{\color{bleu}\nu}}
\def\xx{{\color{bleu}x}}
\def\ggamma{{\color{bleu}\gamma}}
\def\vvarepsilon{{\color{bleu}\varepsilon}}

\def\pphi{{\color{violet}\phi}}

\def\LLambda{{\color{red}\Lambda}}\def\PPsi{{\color{red}\Psi}}
\def\varppi{{\color{red}\varpi}}
\def\ppi{{\color{red}\pi}} \def\pic{\ppi^{\rmc}} 
\def\pif{\ppi^{\rmf}}    

\def\mmu{{\color{red}\mu}} \def\muc{\mmu^{\rmc}}
\def\muf{\mmu^{\rmf}} \def\muI{\mmu^{\rmI}}
\def\cchi{{\color{red}\chi}}
\def\JJ{{\color{red} J}} \def\Jc{\JJ^{\rm c}} \def\Jf{\JJ^{\rm f}}
\def\II{{\color{red} I}} \def\Icc{\II^{\rm cc}}
\def\Iff{\II^{\rm ff}} \def\Icf{\II^{\rm cf}}

\begin{document}
\title
{\bf Effect of entrainment on stress and pulsar glitches\\
in  stratified neutron star crust}

\author { {\bf Nicolas Chamel$^{\star\,\dagger}$, Brandon Carter$^{\star}$}
\\ \hskip 1 cm\\   \\
$$ $^{\star}$ LuTh, Observatoire de Paris, 92195 Meudon, France \\
$$ $^{\dagger}$ Centrum Astronomiczne im M. Kopernika, Warszawa, Polska \\
 $$ }

\date{\it Extended version, 10 December 2005}

\maketitle

\label{firstpage}

\begin{abstract}
The build up of the stress whose relaxation is presumed to account for pulsar
frequency glitches can be attributed to various mechanisms, of which the most
efficient involve differential rotation of the neutron superfluid in the inner
layers of the (magnetically braked) solid crust of a rotating neutron star. In
such a case it is usually supposed that the stress is attributable to pinning
of superfluid vortices to crust nuclei, but it has been suggested that, even 
if the pinning effect is too weak, a comparably large stress might still 
arise just from the deficit of centrifugal buoyancy in the slowed down crust. 
The present work is a re-examination that investigates the way such processes 
may be affected by considerations that were overlooked in the previous work -- 
notably uncertainties about the ``effective'' masses that have to be 
attributed to the ``free'' superfluid neutrons to allow for their entrainment 
by the ionic crust material. Though restricted to a Newtonian formulation, 
this analysis distinguishes more carefully than has been usual between true 
velocities, which are contravariantly vectorial, and so called ``superfluid 
velocities'' that are proportional to momenta, which are essentially
covectorial, a technicality that is important when more than one independent 
current is involved. The results include a Proudman type theorem to
the effect that the superfluid angular velocity must be constant on
slightly deformed Taylor cylinders in the force free case, and it
is shown how to construct a pair of integral constants of the motion
that determine the solution for the pinned case assuming beta equilibrium.

\end{abstract}

\section{Introduction}

This work is intended as a comparison of the qualitative mathematical
consequences of hypotheses of various alternative kinds (particularly
concerning vortex pinning and chemical equilibrium) that have been invoked 
in the construction of differentially rotating neutron star models of
the type developed to acccount for phenomena observed in pulsars. 

\begin{figure}
\centering
\epsfig{figure=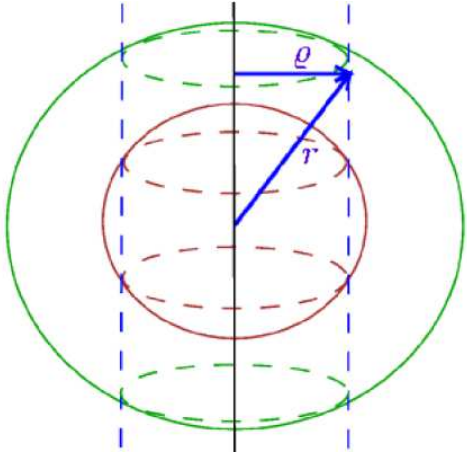, height=6 cm}
\caption{Intersection of rotating neutron star model with a Taylor
cylinder, over which the superfluid angular velocity must be uniform
when the conditions for the Proudman theorem are applicable.}
\label{fig.cylinder}
\end{figure}

It is hoped that this qualitative study will be helpful for the preparation 
of quantitative models designed to improve on those of the simple kind 
constructed by Prix et al \cite{PrixComAnd02}, for which (as for the 
analogously constructed models \citep{PrixNovCom05} in a relativistic 
framework) separately conserved proton and superfluid neutron currents were 
able to deviate from Proudman - Taylor type corotation illustrated on figure
\ref{fig.cylinder}, without giving rise to the
(glitch producing) stresses that are actually anticipated. In earlier
work \citep{Prix99} such stresses were obtained by including allowance
for the effect of vortex pinning, but it has since been remarked 
\citep{CarSedLan} that deviations from corotation would give rise to 
stresses in any case (with or without pinning) when allowance is made for
the consideration that the proton and neutron numbers will not be
separately conserved but will undergo transfusion (by beta processes) 
so as to achieve chemical equilibrium. 

The work of Prix \cite{Prix99} and of Carter et al \cite{CarSedLan} used a 
simplified treatment that neglected the mechanism of relative entrainment 
between the differentially rotating constituents, an effect which had 
already been considered in earlier work (both in a Newtonian framework 
\citep{Mendell91} and also \citep{LanSedCar98} in a relativistic treatment) 
but which had been considered to be only of secondary relevance as a minor
quantitative correction. The effect of entrainment (but not transfusion)
in the liquid core was allowed for in the more recent work of Comer \&
Andersson \cite{ComAnd01} and of Prix et al. \cite{PrixComAnd02}, and it 
has since been recognised \citep{Chamel04} to be particularly important in 
the crust layers that are the primary concern of the present analysis, which 
will consider the (glitch producing) stresses that can be expected to arise 
from the combined effect of entrainment, pinning, transfusion, and also 
nuclear stratification.

Explanations of the glitches of the pulsar period $\period=2\pi/\Om$
observed in rotating neutron stars may be broadly classified into two main 
categories. In what may be described as the ``deformation'' category, the 
discontinuous spin up $\delta\Om$ is attributed to a sudden change of 
the geometric configuration of the matter distribution \citep{Franco00},
whereby the relevant moment of inertia is decreased so that -- for a given 
angular momentum -- the angular velocity $\Omega$ must increase. While
conceivably sufficient to account for cases such as the glitches of order 
$\delta\Om/\Om\simeq 10^{-8}$ observed in the Crab pulsar, such a
deformation mechanism is inadequate for explaining the frequent occurrence 
of the larger gliches $\delta\Om/\Om\simeq 10^{-6}$ observed in Vela 
like pulsars, not to mention the enormous glitch $\delta\Om/\Om
\simeq 1.6\times 10^{-5}$ recently observed in PSR J1806-2125 by 
Hobbs et al. \cite{Hobbs02}.

To account for the frequent occurence of such large glitches, it has long
been generally recognised \citep{Baym69} to be necessary to invoke a
mechanism belonging to what may be referred to as the ``transfer'' category,
involving interaction between two dynamically distinct constituents of the 
star. In such a case the discontinuous increase in the observed angular
velocity $\Om$ is attributed to transfer of angular momentum from a more 
rapidly rotating internal constituent to a more slowly rotating constituent 
that is directly coupled to the outer magnetosphere whose pulsations are 
directly observed. However it is not so clear what actually happens during 
the discontinous process involved. It has been suggested~\citep{Ruderman91} 
that the stress due to pinning of neutron superfluid vortices to the ionic 
crust material may build up elastic stress to the point where the solid 
stucture itself breaks down (as in a terrestrial earthquake). In another 
much studied scenario \citep{Alpar93, Pizzochero97} it is postulated that 
before this can happen there will be a crisis of a different kind whereby a 
threshold is reached at which the vortices in the crust become collectively 
unpinned (whereas in the fluid core it is expected that vortices in the
core will be subject to a strong dissipative drag, which will simulates 
pinning but will not be limited by any threshold). Another possibility
\citep{Jones98, JonesAnd01, Donati03, Donati04} is that pinning in the
crust may be too weak to be effective, but it has been  pointed out
\citep{CarSedLan} that differential rotation might in principle by itself
engender sufficient elastic stress to cause a breakdown of the solid 
structure even in the absence of pinning. 

The main purpose of the present work is not to draw any specific 
conclusions about the relative plausibility of these various postulates in
particular cases, but to consider, generally, how such alternatives will 
affect the predicted outcome, and how estimates of relevant quantities such 
as subcomponent moments inertia and stresses may need to be modified to 
take account of some effects that tended to be neglected in preceding
discussions, including stratification and particularly the entrainment
between the ionic crust material and the ``free'' superfluid neutrons,
which thereby aquire ``effective masses'' that can be defined in various
ways, but that in any case are likely\citep{CCHI, CCH, CCHII} to be large 
enough to make an important difference. 

In order to carry out such a revision, and as a secondary purpose in its 
own right, this work employs and develops a Newtonian formulation of the 
covariant kind \citep{CC03, CC04} whose use would be taken for granted in a
relativistic treatment, in which care is taken to distinguish between true 
velocities, which are contravariantly vectorial, and so called ``superfluid
velocities'' that are proportional to momenta, which are essentially
covectorial. 

Although deeply rooted in the well known principles of Lagrangian and 
Hamiltonian mechanics, the fundamental distinction between velocities and 
their canonical conjugates, namely momenta, has commonly been obscured in 
the context of superfluidity as traditionally formulated \citep{Mendell91}
in a Newtonian framework, for which it has long been customary to use the 
term  ``velocity'' indiscriminately for anything having the physical 
dimensions thereoff, regardless of its actual mathematical role.  The 
distinction is relatively unimportant for a single constituent perfect 
fluid or superfluid such as that of ordinary Helium at zero temperature,
or even for a two contituent fluid in which entrainment is absent
\citep{CarSedLan}, but it becomes important when there are two or more 
independently moving fluids with entrainment, which means that the momenta
need no longer be aligned with their conjugate velocities.

As well as simplifying the algebra in the work presented below, the 
introduction of the  canonically covariant formulation also has the 
advantage of providing a helpful step towards the development of a fully
General relativistic treatment, such as will ultimately be necessary for 
accuracy. Relativistic effects will be particularly important for the 
core, in which however, compared with the crust,
the effect of entrainment is expected to be much more moderate \citep{Sauls89,Comer03}
at least in the outer part for which the basic physics is reasonably
well understood, unlike the inner core for which many exotic
possibilities have been envisages, such that of hyperon
crystallisation \citep{Alonso02}, or of a LOFF phase in a color
superconducting quark condensate \citep{Alford01}.

\section{Two fluid model for stratified neutron star crust}

\subsection{The Lagrangian master function}

The previous treatment referred to above \citep{CarSedLan} used a non 
relativistic Newtonian description in terms of free neutron and confined 
baryon 4-currents $\nf^{\,\nu}$ and $\nc^{\,\nu}$ respectively representing 
a neutron superfluid, and a normal crust component consisting of ionic 
nuclei characterised by a confined baryon number $A_{\rm c}$ and a positive 
charge number $\ZZ$  (in a degenerate electron background) that were 
treated as decoupled but interpenetrating barotropic perfect fluids 
interacting only via vortex pinning and long range gravitational forces,
with the solid rigidity of the crust component taken into account by the 
inclusion of an extra stress force density.

One of the improvements introduced here is to drop the barotropy restriction
for the ionic constituent by allowing its behavior to depend on the
nuclear mass number $A_{\rmc}$ (which will tend to be larger in the deeper 
layers) thus allowing for the possibility of a stratification effect that 
may be important for stabilisation. 

Another improvement that would be needed for a more accurate analysis is of 
course the use of a general relativistic treatment,  of the kind that has 
long been commonly used with a simple perfect fluid description, and that 
has recently been developed for use with a two constituent fluid 
description of the kind needed fore treating the stellar core
\citep{LanSedCar98,AndComLan02}. However for the semilocal modelisation 
needed for the treatment of differential rotation in the crust layers with
which we shall be concerned here, it would appear that corrections from 
general relativity will be much less important (at most a few tens of per 
cent) than those due to the entrainment. A recent program of work 
\citep{CCHI, CCH, CCHII} (using a microphysical analysis combining methods 
from nuclear and solid state physics) on the relative motion of unbound 
neutrons in the inner crust layers (between the neutron drip density about 
10$^{11}$ g/cm$^3$ and nuclear saturation density about 10$^{14}$ g/cm$^3$)
has provided results \citep{Chamel04} indicating that the effect of 
entrainment is likely to be very much stronger than had previously been 
supposed. 

As in the preceding work \citep{CarSedLan}, we shall therefore proceed here
within a non relativistic Newtonian treatment, but instead of using a
decoupled perfect fluid description we shall now include allowance for the
effect of entrainment, using a non-dissipative two constituent description 
of the  generic kind already developed for use in the liquid core
\citep{Mendell91, ComAnd01, PrixComAnd02} where the entrainment corrections 
are relatively moderate, though still potentially important as a cause for 
two-stream instability \citep{AndComPrix03} if the relative velocity 
becomes too large.

For the reasons outlined in the introduction, it is both instructive and
convenient (even though our treatment is non-relativistic) to use a canonical
formalism of the 4-dimensionally covariant kind recently developed
\citep{CC03,CC04}  for the purpose of facilitating the exploitation of the 
variational formulation that is applicable to the non-dissipative limit. In 
this limit there is no resistance to relative motion, so that the only 
direct coupling between the two constituents is via the
entrainment effect \citep{AB76}, whose microscopic origin is the Bragg 
scattering of dripped neutrons by crustal nuclei \citep{CCHI} which is the 
analog in the nuclear context of conduction electron scattering in ordinary 
solids.

It is also convenient to use a chemical constituent label $ _{\rm X}$ that 
runs over two values, namely $ _{\rm X}={\rm f}$ for the superfluid 
constituent and $ _{\rm X}={\rm c}$ for the crust constituent. Each 
constituent will have a baryon current density 4-vector having the form
\be \nn_{_{\rm X}}^{\, \nu}=\nn_{_{\rm X}}\uu_{_{\rm X}}^{\, \nu}\fe
in which the scalars $n_{_{\rm X}}$ are the respective number densities of 
free baryons (only neutrons) and of baryons that are effectively confined 
within the atomic nuclei (namely all the protons, together with a roughly 
equal but somewhat larger number of confined neutrons, whose precise 
specification needs further clarification as discussed below) while the 
vectors $u_{_{\rm X}}^{\, \nu}$ are the corresponding 4-velocities, whose 
components are labelled by a Greek index $\nu$ running over the values 
$0,1,2,3$. The conservation of the total baryon number current
\be \nn_{\rm _B}^{\,\nu}=\nf^{\,\nu}+\nc^{\,\nu}\, ,\fe
is concisely expressible in this 4-dimensional notation as
\be \nabla_{\!\nu} \nn_{\rm _B}^{\,\nu}=0\, .\label{barcons}\fe
In Aristotelian coordinates (representing the usual kind of 3+1 space time 
decomposition with respect to the rest frame of the star)  the 4-velocity 
components will have the form 
$\uu_{_{\rm X}}^{\,_0}=1$, $\uu_{_{\rm X}}^{\, i}=\vv_{_{\rm X}}^{\, i}$ 
where the $\vv_{_{\rm X}}^{\, i}$ are ordinary 3-velocity components, using a 
Latin index $i=1,2,3$. This means that time component of the current density 
is simply equal to the corresponding particle number density 
$\nn_{_{\rm X}}^{_0} = \nn_{_{\rm X}}$,  while the space components are the 
those of the current density 3-vector $\nn_{_{\rm X}}^i$.

In addition to the six independent vectorial components provided by the space 
vectors $\nc^{\, i}$ and $\nf^{\,i}$ and the pair of scalars provided by the 
``confined'' and ``free'' baryon number densities $\nc$ and $\nf$, our model 
will contain a nineth independent variable, namely the scalar provided by the 
nuclear mass number $A_{\rmc}$, or equivalently by the number density $\nI$ 
of nuclei, whose number current will be given by
\be \nI^{\, \nu}= \nI \uu_\rmc^{\,\nu}\, ,\hskip 1 cm
\nI=\nc/A_{\rmc}\, .\label{nucmass}\fe

We shall take the mass per baryon to be  the same (considering the electron 
mass and the proton neutron mass difference to be negligible compared with 
the ordinary proton mass) with fixed value $\mm^{\rmc} = \mm^{\rmf} = \mm$, 
so that the corresponding mass density contributions will simply be given 
by  $\rrho_{_{\rm X}}=\mm\, \nn_{_{\rm X}}$. The total mass density, as 
given explicitly by 
\be\rrho=\mm\, \nn_{\rm _B}=\rrho_\rmf+\rrho_\rmc\, ,\fe
will be important as the Poissonian source of the Newtonian gravitational 
potential scalar, $\pphi$. Within this Newtonian background field, the total 
force balance equation will take the form
\be \nabla_{\!\mu} \TT^\mu_{\ \nu}=\ff_\nu-\rrho\nabla_{\!\nu}\pphi
\, ,\label{fbalance}\fe
in which $\ff_\nu$ is the non gravitational external 4-force density -- if 
any -- that may be acting on the system, and $\TT^\mu_{\ \nu}$ is the stress 
momentum energy density tensor as constructed in terms of the 4-dimensional 
formalism recapitulated below.

The dynamics of system will be governed -- in the manner described in our 
preceding work \citep{CC03, CC04} -- by a Lagrangian master function 
\be\LLambda=\LLambda_{\rm mat}-\pphi\,\rrho\, , \hskip 1 cm
\LLambda_{\rm mat}= \tilde\LLambda_{\rm kin}+\tilde\LLambda_{\rm int}
\label{lampot}\fe
in which $\LLambda_{\rm kin}$ is the ordinary Galilean frame dependent
kinetic energy density, namely
\be\tilde\LLambda_{\rm kin} =\frac{_1}{^2}\left(\rrho_\rmf \vv_\rmf^{\, 2}
+\rrho_\rmc \vv_\rmc^{\, 2}\right)\, ,\fe
and $\tilde\LLambda_{\rm int}$ is some appropriately specified internal
action density contribution that must be Galilean frame independent,
which means that it can depend only on the scalar number densities
$\nI$,  $\nf$, $\nc$, and on the velocity difference
\be  [\vv^i]=\vv_\rmf^{\,i}-\vv_\rmc^{\,i} \fe
so that in terms of an ``entrainment density'' function 
$\bar\rrho_{\rmfc}$ 
and a set of chemical potentials
$\tilde\cchi^\rmf$, $\tilde\cchi^\rmc$, $\tilde\cchi^{\rm_I}$
its generic variation will take the form
\be \delta \tilde\LLambda_{\rm int}=\bar\rrho_{\rmfc} [\vv_i]\,\delta 
[\vv^i] -\tilde\cchi^\rmf\delta \nf-\tilde\cchi^\rmc\delta\nc-\tilde
\cchi^{\rmI}\delta \nI\ ,\label{deLint}\fe
which gives the corresponding formula
\be \delta\LLambda_{\rm mat}=\nf\mmu^\rmf_{\, i}\delta\vv_\rmf^{\, i}
+ \nc\mmu^\rmc_{\, i}\delta\vv_\rmc^{\, i} -\tilde\mmu^\rmf\delta \nf
-\tilde\mmu^\rmc\delta\nc-\tilde\mmu^\rmI\delta \nI\, ,\label{varint}\fe
in which the (Galilean frame dependent) material energy coefficients  
are given in terms of the corresponding (frame independent) chemical
potentials by
\be \tilde\mmu^\rmf=\tilde\cchi^\rmf-\frac{_1}{^2}\mm\, \vv_\rmf^{\, 2}
\, ,\hskip 0.4 cm\tilde\mmu^\rmc=\tilde\cchi^\rmc-\frac{_1}{^2}\mm\,
\vv_\rmc^{\, 2}\, ,\hskip 0.4 cm\tilde\mmu^\rmI=\tilde\cchi^{\rmI}\, ,\fe
and the associated space momenta are given by
\be \muf_{\,i}=\mm\, \vv_{\rmf\, i}+\frac{\bar\rrho_{\rmfc}}{\nf}[\vv_i]
\, ,\hskip 1 cm  \muc_{\,i}=\mm\, \vv_{\rmc \, i}-
\frac{\bar\rrho_{\rmfc}}{\nc}[\vv_i]\, .\label{spacm}\fe

In the low relative velocity regimes that are relevant it will always
be possible to take the internal energy contribution to have the
quadratic form 
\be\tilde\LLambda_{\rm int}=\frac{_1}{^2} \bar\rrho_{\rmfc} [\vv]^2
-\tilde\UU_{\rm ins}\, , \label{indec}\fe
with $\bar\rrho_{\rm fc}$ and the static contribution
$\tilde \UU_{\rm ins}$ specified by appropriate equations of state as
functions only of the scalar densities $\nI$,  $\nc$, $\nf$.
It is to be remarked that to translate our notation to that used by 
Prix et al. \cite{PrixComAnd02} we would need to make the conversions
$\tilde\cchi^\rmf\mapsto\mu^\rmf$, $\tilde\cchi^\rmc\mapsto\mu^\rmc$ and 
$\tilde\LLambda_{\rm int}\mapsto-{\cal E}$ where ${\cal E}$ is what these 
authors have referred to rather loosely as ``internal energy'' density, 
although it is something that should not be confused with the true 
internal energy density $\tilde\UU_{\rm int}$, which will actually  be 
given by
\be \tilde\UU_{\rm int}=\frac{_1}{^2}\bar\rrho_{\rmfc} [\vv]^2+
\UU_{\rm ins}=\bar\rrho_{\rmfc} [\vv]^2-\LLambda_{\rm int}\, .\fe

\subsection{Currents and their conjugate momenta in 4 dimensions}

Instead of treating quantities such as energy and momentum separately, as 
has traditionally been done in a Newtonian framework, it has been found to 
be technically advantageous, as remarked above, to unify them in a
4-dimensional treatment \citep{CC03,CC04} (which was originally inspired by
the corresponding relativistic theory \citep{LanSedCar98}).  The first 
step in this 4 dimensional treatment is to rewrite  the generic variation
(\ref{varint}) in tems of 4-current variations in the form
\be {\rm d}\LLambda_{\rm mat}=\muf_{\,\nu}\, {\rm d} \nf^{\,\nu} + 
\muc_{\,\nu}\,{\rm d}\nc^{\,\nu}-\tilde\mmu{^\rmI}\,{\rm d}\nI\, ,\fe
which provides a direct specification of the material 4-momentum covectors 
$\muf_{\,\nu}$ and $\muc_{\,\nu}$ that play a crucial role in the concise 
canonical formalism to be described below. In terms of the quantities
introduced above their time components will be given by
\be \muf_{\,_0}=-(\tilde\mmu^\rmf+\mmu^\rmf_{\,i}\vv_\rmf^{\,i})\, ,\hskip
1 cm \muc_{\,_0}=-(\tilde\mmu^\rmc+\mmu^\rmc_{\,i}\vv_\rmc^{\,i})\, ,\fe
while their space components will simply be given by (\ref{spacm}).

These material 4-momenta,  $\muf_{\,\nu}$ and  $\muc_{\,\nu}$, give rise 
to corresponding complete ``free'' and ``confined'' baryon momenta, 
$\pif_{\,\nu}$ and $\pic_{\,\nu}$, whose time components include allowance 
for gravitational energy according to the specifications
\be\pif_{\, _0}=\muf_{\, _0}-\mm\phi\, ,\hskip 1 cm
\pic_{\, _0}=\muc_{\, _0}-\mm\phi
\, ,\fe
while their  space components are given simply by 
\be \pif_{\, i}=\muf_{\,i}\, ,\hskip 1 cm
\pic_{\, i}=\muc_{\,i}\, .\fe

It is also useful to introduce the total ionic momentum covector 
$\tilde\ppi^{\rmI}_{\,\nu} , $ which is defined so as to include allowance 
for nuclear energy by
\be\tilde\ppi^\rmI_{\, _0}=A_\rmc\pic_{\, _0} -\tilde\mmu^\rmI
\, ,\hskip 1 cm \tilde\ppi^\rmI_i=A_\rmc\pic_{\, i} \, .\fe
This enables us to express the complete stress energy tensor of the system 
in the very concise  form
\be \TT^\mu_{\ \nu}=\nf^{\,\mu}\pif_{\,\nu}+\nI^{\,\mu}
\tilde\ppi^\rmI_{\,\nu}+\PPsi\,\delta^\mu_\nu\, ,\label{stressen}\fe
in which the material pressure function is given by
\be \PPsi=\LLambda_{\rm mat}-\nf^{\,\nu}\muf_{\,\nu}-
\nc^{\,\nu}\muc_{\,\nu}+\nI\,\tilde\mmu^\rmI\, .\fe
It follows that the force balance equation (\ref{fbalance})
can be rewritten in the form
\be \ff^\rmf_{\,\nu}+\ff^\rmI_{\,\nu}=\ff_\nu\, ,\label{sumf}\fe
in which the 4-force density $ \ff^\rmf_{\,\nu}$ acing on the``free'' 
neutrons is specified in terms of their vorticity tensor 
$\varppi^\rmf{_{\!\mu\nu}}$ by
\be \ff^\rmf_{\,\nu} =\nf^\mu\varppi^\rmf{_{\!\mu\nu}}+\pif_{\,\nu}
\nabla_{\!\mu}\nf^{\,\mu}\, ,\hskip 1 cm
\varppi^\rmf{_{\!\mu\nu}}=2\nabla_{\![\mu}\pif_{\,\nu]}
\, ,\label{3}\fe
while the combined 4-force density acting on the ions and their
 ``confined'' baryons will be specified by the  formula
\be \ff^\rmI_{\,\nu} =2\nI^\mu\nabla_{\![\mu}\tilde\ppi{^\rmI}_{\nu]}
+\nI^\mu\pic_{\,\mu}\nabla_{\!\nu} A_\rmc\, . \fe
This can be decomposed in the form
\be \ff^\rmI_{\,\nu} =\ff^\rmc_{\,\nu} +\nI\nabla_{\!\nu}
\tilde\mmu^\rmI -\delta^{_0}_\nu\nabla_{\!\mu}
(\tilde\mmu^\rmI \nI^{\,\mu})\, ,\label{conforce}\fe
in which the last term contributes only to the time component, while 
the only part that remains in the absence of stratification is the 
contribution of  first term, which is given by an expression of the 
standard form
\be \ff^\rmc_{\,\nu} =2\nc^\mu\nabla_{\![\mu}\pic_{\,\nu]}
+\pic_{\,\nu}\nabla_{\!\mu}\nc^{\,\mu}\, . \fe

\subsection{Equations of state and effective mass}

To complete the specification of the variables introduced in the
previous section it is evidently  necessary to prescribe the particular 
form of the material action density $\Lambda_{\rm mat}$, which is --
for this purpose -- most conveniently decomposible in the form
\be \LLambda_{\rm mat}=\UU_{\rm dyn}-\tilde\UU_{\rm ins}\, ,\fe
wherein  $\tilde\UU_{\rm ins}$ is the static internal energy contribution
introduced in (\ref{indec}), which is specified by an appropriate 
``primary'' equation of state as a function of the three relevant number 
densities $\nf$, $\nc$ $\nI$, while the remaining dynamical energy 
contribution is given \citep{CCHII}   in terms of the ``normal'' (crust) 
velocity 3 vector $\vv_\rmc^i$  and the relative current 3 vector
\be \nn^i=\nf[\vv^i]\, ,
\label{relflow}\fe
by the formula
\be \UU_{\rm dyn}=\rrho\, \vv_{\rmc\, i}\vv_\rmc^{\,i}/2+
\mm \,\nn_i \vv_\rmc^{\,i}+ \nn_i \nn^i/2{\calK}\, ,\label{Udyn}\fe
in which the ${\calK}$ is the ``mobility coefficient''. This quantity  
${\calK}$ is given by a ``secondary'' equation of state as a function of 
the relevant number densities $\nI$, $\nc$, and $\nf$. 

In the analysis that follows a particularly important role will be played 
by the ``free'' neutron 3-momentum covector $\muf_{\, i}$ for which, on  
a local ``mesoscopic'' scale (small compared with the intervortex
separation) the superfluidity property entails the irrotationality
condition $\nabla_{\![i}{\muf}{_{j]}}=0$. It follows from (\ref{Udyn}) 
that the contravariant version (as obtained by contracting with the
ordinary Euclidean space metric $\ggamma^{ij}$) of this  (superfluid)
momentum will be given by an expression of the form
\be \mmu^{\rmf\,i}=\mm_\star \vv_\rmf^{\,i}
+(\mm-\mm_\star)\vv_\rmc^{\, i}\label{supmom}\fe
in terms of an effective mass variable,
\be \mm_\star=\nf/{\calK}\, ,\fe
whose difference from the ordinary baryon mass $m$ is proportional
to the ``entrainment density''
\be \bar\rrho_{\rm fc}=\rrho_\rmf(\mm_\star -\mm)/\mm\, ,\fe
 which -- in much of the inner crust -- now  seems likely \citep{Chamel04}
to be much larger than had previously been supposed. 

The reason why it is convenient for the formulation of the equations of 
state to start with the specification of ${\calK}$ rather than the 
effective mass $\mm_\star$ or the corresponding ``entrainment density'' 
$\bar\rrho_{\rmfc}$ in the decomposition (\ref{indec}) is that  -- 
unlike $\mm_\star$ and $\bar\rrho_{\rmfc}$ -- the ``mobility coefficient'' 
${\calK}$ has the advantage of being physically well defined in a 
manner that does not depend on the choice of the chemical basis that 
will be discussed in Section \ref{chembas}.

Another kind of decomposition that is also independent of the choice of 
chemical basis, and that is instructive for the purpose of comparison with 
the notation used in related work, including the seminal article of
\citep{AB76}, is to introduce the concept of the so called ``superfluid 
3-velocity'' $V^{\rm _S}_i$, which in stricter terminology
should be referered to as ``superfluid momentum per unit mass'', as 
defined by setting
\be  \muf_{\,i}=\mm\, V^{\rm _S}_{\,i}\, \hskip  1 cm
V^{\rm _S i}=\, \vv_\rmc^{\, i}+\frac{\mm_\star}{\mm}[\vv^i]\, .\fe
This enables the dynamical energy density (\ref{Udyn}) to be
rewritten (without a cross term)  in the seductively simple form
\be \UU_{\rm dyn}=\frac{\mm}{2}\left(\nn_{\rm _S}
V^{\rm _S}_{\,i}V^{\rm _S i}+\nn_{\rm _N}
 \vv_{\rmc\, i}\vv_\rmc^{\,i}\right)\, ,\label{nocross}\fe
in which the so called ``superfluid particle number density''
$\nn_{\rm _S}$ and the corresponding``normal particle number density''
$\nn_{\rm _N}$  are defined by
\be \nn_{\rm _S}=\mm{\calK}=\rrho_\rmf/\mm_\star\, \hskip 1 cm
\nn_{\rm _N}=\nn_{\rm _B}- \nn_{\rm _S}\, .\fe

\subsection{Dynamical Equations}

Even after the primary (3 variable) equation of state function
$\tilde\UU_{\rm ins}\{\nI,\nc,\nf\}$ and the secondary (2 variable)
equation of state function ${\calK}\{\nI,\nc\}$ have been
specified, the formulae given so far will merely provide a coherent 
set of definitions, but -- except for the total baryon conservation law 
(\ref{barcons}) -- they contain no dynamical information. To govern the evolution
of the nine independent  variables (namely $\nI$ and the
components of the 4-vectors $\nf^{\,\nu}$ and  $\nc^{\,\nu}$) of the system
we still need eight more conditions, which can be provided by specification 
of the values of the 4-forces  $\ff^\rmf_{\,\nu}$ and $\ff^\rmc_{\,\nu}$. 

In the absence of an external force $\ff_\nu$ in (\ref{sumf}) the sum of the 
contributions  $\ff^\rmf_{\,\nu}$ and $\ff^\rmc_{\,\nu}$ would have to vanish, 
a condition that will automatically be satisfied if these contributions are 
specified just by the variational principle, according to which they should 
each vanish separately,  $\ff^\rmf_{\,\nu}=\ff^\rmc_{\,\nu}=0$.

In a dissipative model o the kind recently developed by \cite{CC04}
(such as might be needed to allow for
the mutual resistivity that would be present if the temperature were too
high for superconductivity) the contributions $\ff^\rmf_{\,\nu}$ and 
$\ff^\rmc_{\,\nu}$ could still add up to zero even though they would no 
longer vanish separately. However we wish to consider more general 
situations in which a non-vanishing total force density $\ff_\nu$ may be 
present for two reasons.  One reason is to allow for the ``secular'' (very 
long timescale) effect of weak magnetic braking of pulsar rotation. However 
the main reason for introducing the external force contribution $\ff_\nu$ in 
the present analysis is to take account of the presence, due to the solidity 
of the crust, of additional stresses that we wish to estimate but for which 
a complete treatment would be beyond the scope of the purely fluid model used 
here.

In conjunction with (\ref{barcons}), strict application of the variation 
principle would entail as consequence that each of the three currents
$\nI^{\,\nu}$, $\nf^{\,\nu}$,  $\nc^{\,\nu}$ should be separately
conserved. This would no doubt be a very good approximation for the purpose 
of treating high frequency oscillation modes. However, in the medium to 
long timescale processes we wish to consider here, whereas it will still be
reasonable to postulate conservation of the number of ionic nuclei, meaning
that we we shall have
\be \nabla_{\!\nu}\nI^{\,\nu}=0\, ,\label{ioncon}\fe
on the other hand it may be realistic to allow for the possibility that -- 
due to beta processes whereby neutrons are transformed into neutrons
or vice versa -- the ``free'' and ``confined'' baryon currents may not be
separately conserved, but will evolve in such a way as to diminish the
magnitude of the relevant chemical affinity\citep{CC04}, $\tilde{\calA}$ 
say, meaning the chemical potential difference between the two constituents, 
as measured in the relevant  ``normal'' (non superfluid) rest frame, namely 
that of the ionic lattice with unit velocity 4-vector $\uu_{\rmc}^{\,\nu}$, 
so that it will be given by
\be \tilde{\calA}=\uu_{\rmc}^{\,\nu}[\mmu_\nu]\, ,\hskip 1 cm
[\mmu_\nu]=\muf_{\,\nu}-\muc_{\,\nu}\, .\label{chem}\fe

As discussed in the treatment of the analogous relativistic model
\citep{LanSedCar98} (with $\nI$ and $\muI$ respectively replaced by
entropy density and temperature) a non-dissipative model that is
self contained -- so that total force $\ff_\nu$ vanishes --
will be obtainable in a manner that is compatible either with the
short term conservation conditions
$\nabla_{\!\nu}\nf^{\,\nu}=\nabla_{\!\nu}\nc^{\,\nu}=0$, or alternatively
with the long term equilibrium condition $\tilde{\calA}=0$ that is more
relevant here, provided the three force acting on the superfluid
neutron current is specified in such a way that
\be \nf^{\,\mu}\varppi^\rmf{_{\!\mu\nu}}
 \nc^{\,\nu}=0\, .   \label{nodis} \fe

\subsection{Superfluidity condition}

On a mesoscopic scale (large compared with the ionic spacing but small
compared with the distance between quantised vortices) the superfluidity
property of the ``free'' neutrons entails that their momentum covector 
$\pif_{\,\nu}$ must be locally proportional to the gradient of a quantum 
phase scalar, and hence that their vorticity 2-form must vanish,
$\varppi^\rmf{_{\!\mu\nu}}=0$ , which automatically ensures that the
condition (\ref{nodis}) will be satisfied.

\begin{figure}
\centering
\epsfig{figure=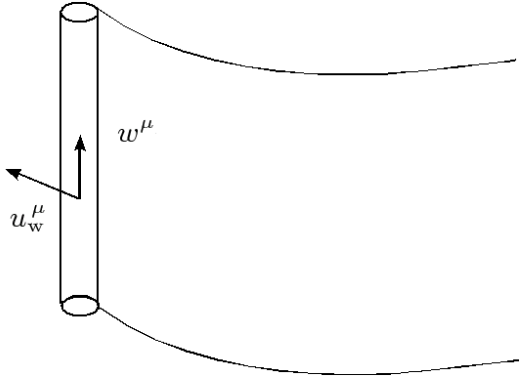, height=5 cm}
\caption{Sketch of the 2-surface swept out by a quantised vortex line.}
\label{fig.vortex}
\end{figure}

What we are concerned with here however is the macroscopic level -- meaning 
lengthscales large compared with the intervortex separation -- at which 
there will be a non-vanishing vorticity 2-form, and a corresponding
spacelike vorticity vector
\be \ww^{\lambda}=\frac{_1}{^2}\varepsilon^{\lambda\mu\nu}
\varppi^{\rmf}_{\,\mu\nu}\, ,\label{wvector}\fe
whose components in Aristotelian coordinates are simply $\ww^0=0$
and $\ww^i=\varepsilon^{ijk}\nabla_j \vv_{\rmf k}$,
representing the mean density of circulation around the quantised vortices.
The superfluidity property does however entail an algebraic restriction to
the effect that we must have
\be \ww^\mu\varppi^{\rmf}_{\,\mu\nu}=0\, ,\fe
a condition need not be satisfied for a more general fluid motion (such as
would be possible for the free neutrons in a very young neutron star that 
has not yet dropped below the temperature of the BCS pairing transition).
This restriction expresses the requirement that,
instead of having matrix
rank 4 as in the normal case, the antisymmetric (so necessarily even 
ranked) vorticity tensor $\varppi^{\rmf}_{\,\mu\nu}$ should have its rank 
reduced to 2 in the superfluid case, a condition that is mathematically 
necessary for it to  be tangential to a set of 2-surfaces which in this 
instance are swept out by the quantised vortex lines as illustrated on
figure \ref{fig.vortex}. This implies the
existence of vorticity transport 4-vector field $\uu_{\rm w}^{\,\nu}$ 
subject to the usual time normalisation condition $\uu_{\rm w}^{\,_0}=1$ 
characterised by the defining property
\be \uu_{\rm w}^{\,\mu}\varppi^{\rmf}_{\,\mu\nu}=0\,\label{vortrans}\fe
which does not fix it completely but evidently allows a freedom of adjustment
by addition of a spacelike vector field aligned with the direction of
the vortex lines as given by $\ww^\nu$.

The supposition that the vortices will be ``pinned'' in the sense of being 
transported with the ionic lattice is evidently expressible by adopting a
dynamic equation of the form (\ref{vortrans}) with
\be \uu_{\rm w}^{\,\mu}=\uu_{\rmc}^{\,\mu}\, \label{pinning}\fe
which still automatically ensures that the condition (\ref{nodis}) will
be satisfied, but which will generically entail the involvement of a 
non-vanishing force density that will be attributable to the Magnus effect 
as discussed in the appendix. This condition (\ref{nodis}) particularly  
plausible as a description of evolution on very long timescales, for which
the beta equilibrium condition,
\be \tilde{\calA}=0\, ,\label{beteq} \fe
will also be appropriate.

For evolution on the more moderate timescales that will be relevant
for many astrophysical processes the situation is not so clear.
An obvious -- though not necessarily realistic -- alternative
way of satisfying (\ref{nodis}) is to apply the strict variation
principle to the effect that $\ff^\rmf_{\,\nu}$ should vanish, which
instead of (\ref{pinning}) as well as the separate ``free'' neutron conservation condition
$\nabla_{\!\nu}\nf^{\,\nu}=0$ also provides
a dynamical equation of the form (\ref{vortrans}) but with
\be \uu_{\rm w}^{\,\mu}=\uu_{\rmf}^{\,\mu}\, ,\label{nopinning}\fe
and which instead of (\ref{beteq}) provides a separate conservation 
law 
\be \nabla_{\!\nu}\nf^{\,\nu}=0 \, .\label{freecon}\fe
However the realism of such an alternative is questionable on the basis of
microphysical considerations which suggest that both the vortex slipping 
implied by (\ref{nopinning}) and the beta disequilibrium entailed by 
(\ref{freecon}) would in practise give rise to dissipative processes that 
would require a more elaborate treatment. This issue is hinted at, on a 
purely macroscopic level, by the consideration that the actual physical 
meaning of the simple set of non dissipative evolution equations provided by 
(\ref{nopinning}) and (\ref{freecon}) is open to question, due to its gauge 
dependence in the sense discussed in the next section.

\subsection{Chemical basis dependence}
\label{chembas}

A problem with the force free postulate, namely the use of (\ref{nopinning})
in conjunction with (\ref{freecon}) is that (unlike the use of 
(\ref{pinning}) in conjunction with (\ref{beteq}) which will be
appropriate for longer timescale processes) its physical meaning depends
on how many of the neutrons are considered to be ``free''. To be more
explicit, it depends on the choice of the dimensionless parameter $a_\rmc$ 
in the chemical base transformation
\be \nc^{\,\nu}=a_\rmc\,\nn_\rmp^{\,\nu}\hskip 1 cm \nf^{\,\nu}=
\nn_\rmn^{\,\nu}+(1-a_\rmc) \nn_\rmp^{\,\nu}\label{cgauge}\, ,\fe
whereby the ``confined'' and ``free'' baryon currents are specified in 
terms of the more obviously well defined chemical basis constituted by the 
proton current $\nn_\rmp^{\,\nu}$ and the total (``confined'' plus ``free'') 
neutron current $\nn_\rmn^{\,\nu}$.

Obvious simple possibilities for the choice of such a ``chemical gauge'' 
include the ``comprehensive'' gauge, $a_\rmc$ in which all the neutrons are 
counted as ``free'', the ``pairing gauge'' $a_\rmc=2$ in which the most 
strongly bound neutron states, namely those paired with proton states, are 
counted as ``confined'', but all the others are considered to be free.  
A more intuitively ``natural'' -- but more fuzzily defined --
possibility would be to reserve the qualification ``free'' for neutrons that 
are actually located outside the -- somewhat blurred -- boundaries of the 
nuclei that contain the protons.  However  what really matters is not 
location in ordinary space but in phase space: the ``operational'' 
criterion for a neutron state to be effectively ``free'' is that it should
have sufficient energy to overcome the barriers separating the ionic
potential wells either classically or by quantum tunnelling within the 
relevant timescale. 

On this ``operational'' basis it is clear that all the neutrons will be 
effectively ``free'' in the fluid core and that none of them will be
effectively ``free'' in the outer crust below the ``drip'' density threshold,
while at densities just above this threshold there will be a clear cut
critical energy above which the neutrons will be able to travel
over ionic separation distances on a microscopically short timescale and
below which their wave functions will be exponentially suppressed outside
the ionic wells so that they will be able to tunnell only on
cosmologically long timescales. However in the deeper and denser 
layers of the crust (which are likely to be particularly important for 
glitch processes) this ``operational'' distinction will no longer be 
so clear cut, because the ionic wells will get too close for the exponential 
suppression outside to be fully effective, so that there will be marginally 
bound states with intermediate penetration timescales that are 
macroscopically long but cosmologicaly short. This means that the
most appropriate way of precisely defining an ``operational'' basis
will depend on the timescales involved in the astrophysical
context under consideration.

In view of the debatability of this question about which basis may be most
appropriate, it is important to observe that the counting convention
used to define the chemical gauge parameter $a_\rmc$ in (\ref{cgauge}) 
makes no difference at all to the specification of the ``normal'' velocity 
$\vv_\rmc^{\,i}$ and the relative current $\nn^i$ defined by (\ref{relflow}) 
and hence that it has no effect on any of the terms in the original 
specification (\ref{Udyn}) for the dynamical energy density. It is no 
less important to observe that uncertainties about the appropriate 
choice for $a_\rmc$ make no difference to the specification of the ``free''
momentum covector $\muf_{\,\nu}$, nor in consequence, to the vorticity 
form $\varppi^\rmf_{\,\mu\nu}$ and the so called ``superfluid 3-velocity'' 
$V^{\rm _S}_{\,i}$, which means that the quantities appearing in
the rewritten formula (\ref{nocross}) will also be unaffected.

The rule chosen for the specification of $a_\rmc$ and hence of $\nf$ will 
evidently affect the individual terms in the formula (\ref{stressen}) for
the stress energy tensor, but it can be seen that the total, 
$\TT^\mu_{\ \nu}$,  will be unaffected by the value of $a_\rmc$ so long as 
it is held fixed, and  even (though this is less obvious) if it is allowed 
to vary, so long  as this ratio,
\be a_\rmc=\nc/\nn_\rmp=A_\rmc \nn_{\rm I}/\nn_\rmp \, ,\fe
is postulated to depend only on the nuclear charge number $\ZZ$, which is
itself definable as the ratio 
\be \ZZ=\nn_\rmp/\nn_{\rm _I}\, .\fe
It thus follows from (\ref{fbalance}) that provided $a_\rmc$ (or 
equivalently $A_\rmc$) is chosen as some function just of the single
variable $\ZZ$, although the choice of this function will indeed affect 
the specification of the separate force density contributions
 $\ff^\rmf_{\,\nu}$ and $\ff^\rmc_{\,\nu}$, nevertheless it will not have 
any effect on their total (\ref{sumf}). This means that it will not have
any effect on the complete set of evolution equations given by specifying 
the total external force $\ff_\nu$ (e.g. by postulating that it should 
vanish) in conjunction with the ionic number conservation law (\ref{ioncon}) 
together with the (long timescale) pinning and beta equilibrium conditions
(\ref{pinning}) and (\ref{beteq}). However if (\ref{beteq}) is replaced 
by (\ref{freecon}), then the physical specification of the system would 
depend on the ansatz for the function $a_\rmc\{\ZZ\}$, which for very short 
timescales could realistically be taken to be simply the ``comprehensive'' 
value $a_\rmc=1$,  corresponding just  to separate conservation of protons
and neutrons. Similarly if (\ref{pinning}) is replaced by   
(\ref{nopinning}) the result would again depend on the ansatz for 
$a_\rmc\{\ZZ\}$, but in this case the appropriate choice is a question 
needing further investigation at a microscopic level.

The strategy of the following work is not to adopt any particular
choice of gauge but to see how far it is possible to draw general 
conclusions that will be valid whatever the choice of the function 
$a_\rmc\{\ZZ\}$ and whatever the postulate adopted for the specification 
of the vortex drift velocity $\uu_{\rm w}^{\,\mu}$ in (\ref{vortrans}). 

\section{Axisymmetric configurations} 

\subsection{Angular momentum convection and the lag formula}

\begin{figure}
\centering
\epsfig{figure=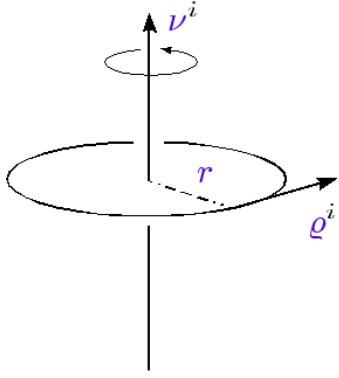, height=5 cm}
\caption{Axisymmetry generator $\hh^i$, whose magnitude is equal to the 
cylindrical radius $\hh$ and the unit vector $\nnu^i$ directed along the 
rotation axis.}
\label{fig.axisym}
\end{figure}

As in the preceding work \cite{CarSedLan} on this problem, we now 
restrict our attention to configurations that are axisymmetric in the 
sense of being invariant with respect to the action of a rotation 
symmetry generator $\hh^\nu$ of the usual kind, meaning one that is 
spacelike, $\hh^{_0}=0$, with space components $\hh^i$ given in terms of
the unit vector $\nnu^i$ along the relevant symetry axis, with respect
to Cartesian coordinates $\rr^i$ centered on the axis, by an expression
of the form $\hh^i=\vvarepsilon^{ijk}\nnu_j \rr_k$ as shown on figure
\ref{fig.axisym}. This means that this symmetry generator will have a scalar
magnitude $\hh$ that can be interpreted as a cylindrical radial coordinate
(such as was denoted by $\varpi$ in the preceding work by Carter et al. 
\cite{CarSedLan}) so that it will be given in terms of the relevant angle of 
latitude $\ttheta$ by $\hh=\rr \,{\rm cos}\, \ttheta$.

Invariance of a field with respect to such an action is expressible in any
(Cartesian or other) coordinate system as the vanishing of its Lie derivative
with respect to the generator. In the particular case of the superfluid
4-momentum covector $\pif_{\,\nu}$ the axisymmetry condition will
therefore be expressible as
\be \hh^\nu\nabla_{\!\nu}\pif_{\,\mu}+
 \pif_{\,\nu}\nabla_{\!\mu}\hh^\nu=0\, .\label{axisym}\fe

In terms of the angular momentum per free (superfluid) particle, as defined 
(independently of any choice of the chemical gauge parameter $a_\rmc$) by
\be \elh=\hh^\nu\pif_{\,\nu}\, ,\fe
(so that on a microscopic scale $\elh$ will be quantised as a half integer
multiple of the Dirac-Planck constant $\hhbar$) the axisymmetry condition
(\ref{axisym}) can evidently be rewritten as
\be \nabla_{\!\mu}\elh=\varppi^{\rmf}_{\,\mu\nu}\hh^\nu\, .\fe
We can use this in conjunction with \rf{vortrans}\fr to obtain a generalised 
Bernoulli type \citep{CC04} conservation law to the effect that this 
superfluid angular momentum scalar $\elh$ will be convected by (in the sense
of remaining constant along) the vortex flux trajectories:
\be \uu_{\rm w}^{\,\nu}\nabla_{\!\nu}\elh=0\, . \label{consang}\fe

We are concerned here with configurations that differ only very slightly from
a state of circularity, meaning a state in which the 3-velocities are all
aligned with the axisymmetry generator $\hh^i$, so that we can write
\be \vv_{\rm_X}^i=\Om_{\rm_X}\hh^i+\vv_{\rm_X\perp}^{\ i}\, ,
\hskip 1 cm \hh_i\,\vv_{\rm_X\perp}^{\ i}=0\, ,\label{convel}\fe
where $\Om_{\rm_X}$ is the relevant angular velocity, and the residual 
non circular velocity contribution $\vv_{\rm_X\perp}^{\ i}$ is supposed to be 
very small except perhaps in very brief intervals during a glitch. The non 
circular part, if any, will evidently make no contribution to the 
superfluid angular momentum scalar, which, according to \rf{superm}\fr
will be given simply by
\be \elh=\hh^2\Big(\mm_\star\Of+(\mm-\mm_\star)\Oc\Big)\, .\label{fang}\fe

Since the axisymmetry evidently implies $\hh^i\nabla_{\!i}\elh=0$, the 
conservation law \rf{consang}\fr can be rewritten in terms of the vortex 
drift 3-velocity $\vv_{\rm w}^{\, i}$ as a slow variation rule of the form
\be \dot\elh=-\vv_{\rm w\perp}^{\ i}\nabla_{\!i}\elh\, ,\label{slowv}\fe
with the usual convention that a dot denotes partial differentiation with 
respect to time at a fixed space position.

In the rotating neutron star configurations under consideration it will be 
justifiable to use a small perturbation approximation, not just for the 
non-circular velocity contributions, but also for deviations of the circular 
part from rigid rotation with a uniform fixed angular velocity value, 
$\bar\Om$ say. This means that for each constituent we can write
\be \Om_{\rm_X}=\bar\Om+\Delta\Om_{\rm_X}\fe
where, like $v_{\rm w\perp}^{\ i}$ the deviation $\Delta\Om_{\rm_X}$ is to 
be considered as relatively small. It follows that to first perturbative 
order we can rewrite the slow variation law (\ref{slowv}) as
\be \dot\elh=-\mm\bar\Om\, \vv_{\rm w\perp}^{\ i}\nabla_{\!i}\hh^2
\, .\label{slow}\fe
Using square brackets for differences between the constituents
according to the convention
\be [\Om]=\Of-\Oc\, ,\hskip 1 cm [\dot\Om]=\dot\Of-\dot\Oc
\, ,\label{sqbrac}\fe
we can write \rf{slow}\fr more explicitly as
\be \mm_\star[\dot\Omega] +\mm\dot\Oc
=-\frac{2 \mm\bar\Om}{\hh}\vv_{\rm w\perp}^{\ i}\nabla_{\!i}\hh \, .\fe
It follows that, after a finite time interval during which the vortices 
have undergone a small cylindrical radial displacement, $\Delta \hh$ 
say, the ensuing (local) change in the angular frequency lag will be 
given in terms of the corresponding  (uniform) change $\Delta\Om$ in the 
observable rotation frequency, namely that of the solid constituent 
$\Om=\Oc$ by the remarkably simple formula
\be \frac{\Delta[\Om]}{\Om}=-\frac{\mm}{\mm_\star}\left(\frac
{\Delta\Om}{\Omega}+2\frac{\Delta \hh}{\hh}\right)\, .\label{lag}\fe

In the idealised case of pinning to an absolutely rigid solid the 
displacement $\Delta \hh$ would of course vanish, so that $\Delta[\Om]$ 
would be approximately determined as a function just of the spherical 
radial coordinate $\rr$ by a proportionality relation of the form  
$\Delta[\Om]\propto\mm/\mm_\star$, but in practice, even for perfect
pinning, there will be deviations from this as the crustal rigidity 
modulus is expected \citep{Ruderman91} to be rather low (the solid 
structure will be effectively ``squashy''). On the other hand if it is 
supposed that pinning is entirely ineffective (since the vortices
may drift by either thermal \citep{Alpar84} or quantum fluctuations 
\citep{Link93}) then the generalised Proudman theorem derived below in 
Subsection \ref{Proud} tels us that the angular velocity difference
$[\Om]$ -- and hence also its variation  $\Delta[\Om]$ -- will be a 
function not of $\rr$ but of the cylindrical coordinate $\hh=\rr\, 
{\rm cos}\,\ttheta$, so in that case (to allow for non-uniformity of  
$\mm_\star$) the contribution from the displacement term in (\ref{lag})  
would have to be relatively significant. 

\subsection{Simplified global model}

For the purpose of drawing conclusions from available empirical data
such as that of Lyne et al. \cite{Lyne00}, rather than with the locally 
well defined but observationally inaccessible quantities involved in a 
formula such as (\ref{lag}) it can be more instructive in practice to work 
with corresponding crudely defined averages, using a total angular 
momentum decomposition of the form 
\be \JJ=\Jf+\Jc\fe
in terms of ``free'' and ``confined'' contributions of the form
\be \JJ^{\rm _X}=\int\, \nn_{\rm _X}\mmu^{\rm _X}_{\ i}\,\hh^i\, 
{\rm d}^3 \rr\, ,\fe
with the ``free'' bayon momentum $\muf_{\, i}$ given by (\ref{supmom})
and with the  ``confined'' baryon momentum given by tha analogous formula
\be\mmu^{{\rmc}\,i}=\mm \vv_{\rmc}^{\,i}- \mm^{\rmc}_\star\, 
[\vv^i]\, ,\hskip 1 cm \, ,\label{superm}\fe
in which the effective mass function $\mm^{\rmc}_\star$ for the ``confined'' 
baryons is determined by its ``free'' baryon analogue,analogue 
$\mm^\rmf_\star=\mm_\star$, via the symmetric relation
\be \nc(\mm^{\rmc}_\star-\mm)=\nf(\mm_\star-\mm)\, .\label{entrm}\fe
Using an approximation in which $\Of$ is taken to be roughly uniform, as
$\Oc (=\Omega)$ must be in any case, these contributions will be given 
roughly by 
\be \Jf\simeq\Iff\Of+\Icf\Oc\, ,\hskip 1 cm \Jc\simeq\Icc\Oc+\Icf\Of
\, ,\label{angdec}\fe
in terms of coefficients that combine to give the respective
moments of inertia of the confined and free parts in the form
\be \II^{\rm f}=\Iff+\Icf\, ,\hskip 1 cm \II^{\rm c}=\Icc+\Icf
\,  .\fe
In accordance with  \rf{superm}\fr the separate
coefficients will be given, in terms of the cylindrical
radius $h$,  as the volume integrals 
\be \Icc=\int \mm^{\rmc}_\star\,\nc\,\hh^2\, {\rm d}^3 \rr\, ,\hskip
1 cm \Iff=\int \mm_\star\,\nf\,\hh^2\, {\rm d}^3 \rr\, ,\label{Inc}\fe
while, in view of the equivalence \rf{entrm}\fr, the entrainment
coefficient will be given by
\be \Icf=\int (\mm-\mm^{\rmc}_\star)\,\nc\,\hh^2\, {\rm d}^3 \rr=
\int(\mm-\mm_\star)\,\nf\,\hh^2\, {\rm d}^3 \rr\, ,\fe
so the subtotals will be given by expressions of the familiar form
\be \II^{\rm c}=\mm\int \nc\,\hh^2\, {\rm d}^3 \rr\, ,\ \hskip
1 cm \II^{\rm f}=\mm \int \nf\,\hh^2\, {\rm d}^3 \rr\, ,\fe
which combine to give the complete stellar moment of inertia as
\be \II=\II^{\rm c}+\II^{\rm f}=\int \rrho\,\hh^2\, {\rm d}^3 \rr
\, .\fe

In a simplified treatment it may be supposed that these moment of 
inertia coefficients remain constant (i.e. that the effects of a 
conceivable  radial displacement $\Delta\, \hh$ in (\ref{lag}) can 
be ignored) and that the superfluid contribution to the angular
momentum is conserved during the interglitch period, 
$\Delta\Jf\simeq 0$, which by (\ref{angdec}) evidently implies 
\be \Delta\Of\simeq -\frac{\Icf}{\Iff}\,\Delta\Om\, .\fe
It may also be supposed that the total angular momentum variation 
is negligible, $\delta \JJ=0$ during the very short time interval
in which the star undergoes a discontinuous glitch transformation
involving an angular momentum transfer given by 
$\delta\Jf\simeq-\delta\Jc$ so that the corresponding angular 
velocity discontinuities will be related by
\be \delta \Of\simeq- \frac{\II^{\rmc}}{\II^{\rmf}}\,
\delta \Om\, .\fe
To avoid a long term build up of too large a deviation of the 
superfluid angular velocity $\Of$ from the externally observable 
value, $\Oc=\Om$, the averages over many glitches of their 
total (continuous plus discontinuous) change per glitch should be the 
same,
\be \langle\Delta\Of+\delta\Of\rangle\simeq \langle\Delta\Om
+\delta\Om\rangle\, ,\fe
a condition that is equivalently expessible using the notation of
\rf{sqbrac}\fr as
\be \langle\delta[\Om]\rangle\simeq-\langle\Delta[\Om]\rangle\, .\fe 
The immediately preceding relations can be used to rewrite this
condition in the form
\be \frac{\langle\delta\Om\rangle}{\langle\Delta\Om\rangle}
\simeq- \frac{(\II^{\rmf})^2}{\Iff\,\II} \, ,\fe
in which the left hand side contains only quantities that are
directly observable, with values that have been typically
found \citep{Lyne00} to be given in order of magnitude by 
${\langle\delta\Om\rangle}/{\langle\Delta\Om\rangle}
\approx -10^{-2}$
in the case of Vela type pulsars. This magnitude has commonly been 
used as a basis for estimating the ratio $\II^{\rm f}/\II$ of 
the moment of inertia of the superfluid component involved to that of 
the whole star, on the basis of the usual supposition that there is 
no need to distinguish between the coefficients $\II^{\rmf}$ 
and $\Iff$. However it can be seen from \rf{Inc}\fr that the
relation between these coefficients will actually be expressible in 
terms of a suitably weighted mean value $\bar \mm_\star$ of
$\mm_\star$ by the formula 
\be \Iff\simeq\frac{\bar \mm_\star} {\mm} \II^{\rmf}\, ,\fe
in which, according to the recent work \citep{Chamel04} referred to
above, the ratio $\bar \mm_\star/\mm$ is likely to be
substantially larger than unity, which leads to a correspondingly 
augmented estimate,
\be \frac{\II^{\rmf}}{\II}\simeq-\frac{\bar \mm_\star}{\mm}\frac
{\langle\delta\Om\rangle}{\langle\Delta\Om\rangle}\, ,\label{momIm}\fe
for the relative value of the moment of inertia $\II^{\rm f}$
of the relevant superfluid component as compared with the total
moment of inertia $\II$ of the neutron star.

\section{Stationary circularly symmetric configurations.}

\subsection{Total stress force in the general case}

We now investigate the equilibrium conditions that need to be satisfied
in the short and medium term when we neglect the slow secular evolution 
effects considered in the preceding section and consider the system to 
to be in a state that is exactly stationary, and with flow that is
strictly circular in the sense that convective velocity contributions
$\vv_{\rm_X\perp}^{\ i}$ in (\ref{convel}) are absent, so that we 
simply have
\be \vv_\rmf^{\,i}=\Of\, \hh^i\, ,\hskip 1 cm
\vv_\rmc^{\,i}=\Of\,  \hh^i\, .\label{circularity}\fe
For a single-fluid model in such a non convective state
the well known Proudman theorem tells us that the angular velocity
must be uniform over each ``Taylor cylinder'' (as characterised
by a fixed value of $\hh$) for the case of a perfect fluid that is
barotropic, meaning that there is no stratification. It will be shown
in the present section how this result can be generalised to
the two-fluid context under consideration here.

Using the circularity conditions (\ref{circularity}) in conjunction
with the axisymmetry conditions exemplified by (\ref{axisym}), and 
assuming that the solidity property of the crust will ensure that the 
``normal'' flow velocity of the ``confined'' current in such a 
stationary configuration will satisfy the rigidity condition,
meaning that $\Oc$ is uniform with the externally
observed angular velocity $\Om$,
\be \nabla_{\! i\,}\Oc=0\, ,\hskip 1 cm \Oc=\Om\, ,\fe
it can be seen that the space parts of the force density formulae 
(\ref{3}) and (\ref{conforce}) will be expressible simply as
\be \ff^\rmf_{\, i} =\nf \left(\nabla_{\! i\,}{\cC}_\rmf
 +\elh\, \nabla_{\! i\,}\Of \right) \, , \label{forcef}\fe
\be \ff^\rmI_{\, i} =\nc \nabla_{\! i\,}{\cC}_\rmc
+\nI\nabla_{\! i\,} \tilde\mmu{^\rmI} \, ,\label{forcec}\fe
using the notation of  Prix et al. \cite{PrixComAnd02} for the comoving
particle energies, which are concisely specifiable for the respective 
``free'' and ``confined''  constituents as the corresponding
diagonal components 
\be {\cC}_\rmf={\calE}^\rmf_{\ \rmf}\, ,\hskip 1 cm {\cC}_\rmc
={\calE}^\rmc_{\ \rmc}\, , \fe
of the (non-symmetric) energy matrix defined by the formula
\be {\calE}^{\rm_X}_{\,\rm _Y}=-\ppi^{\rm X}_{\,\nu} 
\uu_{\rm Y}^{\,\nu} \, .\label{ematrix}\fe
This is another example of a formula that becomes more complicated
and less heuristically meaningful when instead of using 4 dimensional
notation it is written out using the traditional 3 dimensional notation,
in which for example the comoving particle energy for the ``free'' 
constituent will be given by the formula
\be {\cC}_\rmf= \mm\phi+\tilde\cchi{^\rmf}-\frac{_1}{^2} 
\mm\, \vv_\rmf^{\,2}\, ,\fe
in which, by (\ref{deLint}) and (\ref{indec}), the relevant chemical  
potential will be given by
\be\tilde\cchi{^\rmf}=\partial \tilde\UU_{\rm ins}/\partial \nf-\frac{_1}{^2}
[\vv]^2\partial\bar\rrho_{\rmfc}/\partial \nf\, , \fe
with $\partial\bar\rrho_{\rmfc}/\partial \nf= \nf\, 
\partial \mm_\star/\partial\nf +\mm_\star-\mm$.

Without going into such details, it can be seen that the gravitational 
contributions to these comoving particle energies will cancel out in their 
difference, which will be given in terms of the chemical affinity 
(\ref{chem}) and the angular velocity difference $[\Om]
=\Of-\Om$ by 
\be {\cC}_\rmc-{\cC}_\rmf=\tilde{\calA}+\elh\,[\Om]
\, .\label{endif}\fe

If the crust were purely fluid, the force density $\ff_i^{\rmc}$ would 
just have to be equal in magnitude and opposite in sign to the force 
density acting on the superfluid, namely $-\ff_i^{\rmf}$. Detailed 
evaluation of the effects of solidity would require the use of a more 
elaborate model incorporating elastic rigidity, but as in the preceding 
work by Carter et al. \cite{CarSedLan} we can allow for the effect of 
finite rigidity by introducing a non vanishing stress force density 
$\ff_{i}$ representing the discrepancy according to the definition 
(\ref{sumf}), which can be rewritten in the form
\be \frac{1}{\nn_{\rmc}}\left(\ff_{i}-\frac{\rrho}{\rrho_{\rmf}}
\ff^{\rmf}_{\,i}\right)= \frac{\ff^{\rmI}_{\,i}}{\nn_{\rmc}}-
\frac{\ff^{\rmf}_{\,i}}{\nn_{\rmf}}\, ,\label{diff}\fe
in order to obtain a difference on the right hand side that is
particularly amenable to simplification using (\ref{endif}). In terms of 
the chemical disequilibrium force density term defined by
\be \ff^\chi_{\, i}=\nc\nabla_{\! i\, }\tilde{\calA}+\nI
\nabla_{\! i}\tilde\mmu^{\rmI}\, ,\label{chemdis}\fe
it can be seen that the formulae (\ref{forcef}) and (\ref{forcec})
lead to an expression of the remarkably simple form
\be \frac{\ff^{\rmI}_{\,i}}{\nc}-
 \frac{\ff^{\rmf}_{\,i}}{\nf}=\frac{\ff^\chi_{\, i}}{\nc}
+[\Om]\nabla_{\!i}\elh\, .\fe
We thus obtain an expression of the same form
\be \ff_{i}=\frac{\rrho}{\rrho_{\rmf}}\ff^{\rmf}_{\,i}+
\ff^{\chi}_{\, i}+\ff^{\rmb}_{\,i}\, ,\label{stress}\fe
as that previously derived by Carter et al. \cite{CarSedLan} in a 
simplified treatment in which the effects of entrainment and stratification 
were not included, with the force density attributable to centrifugal 
buoyancy deficit given by the same formula as previously, namely
\be \ff^{\rmb}_{\, i}=\nc[\Om]\nabla_{\!i}{\elh}\, .\label{boy}\fe

\subsection{Generalised Proudman theorem for force-free superfluid case}
\label{Proud}

Leaving aside the pinned case for treatment in the next subsection,
let us restrict our attention in the remainder of the present subsection
to the case in which the neutron  superfluid flow is postulated to
be force-free, which means that we have
\be \ff^{\rmf}_{\,i}=0 \, .\fe
In this case the chemical and buoyancy contributions on the right of 
\rf{stress}\fr will be the only sources for the stress force, which in 
this case will be given simply by 
\be \ff_{i}=\ff^{\chi}_{\,i}+\ff^{\rmb}_{\,i}\, .
\label{unpineq}\fe
Under these circumstances (\ref{forcef}) reduces to the form
\be \nabla_{\! i\,}{\cC}_\rmf
 +\elh\, \nabla_{\! i\,}[\Om]=0\, ,\label{Proudman}\fe
in which by (\ref{fang}) we have
\be \elh=\hh^2\Big(\mm\,\Om +\mm_\star[\Om]\Big)\, .\label{fangm}\fe

Taking the exterior derivative of equation (\ref{Proudman})
implies that $\vvarepsilon^{ijk}(\nabla_{\!i}\elh) \nabla_{\! j\,}[\Om]$ must vanish
so that $\nabla_{\! i\,}[\Om]$ has to lie in the direction of
$\nabla_{\!i}\elh$ which is approximately that of $\nabla_{\!i}\hh$.
Consequently the condition (\ref{Proudman}) can be seen to provide us with a Proudman
type theorem telling us that the  energy parameter ${\cC}_\rmf$ and the
angular velocity difference $[\Om]$ must both be uniform over each of the 
deformed Tayor cylinders characterised by a fixed value of the angular 
momentum per ``free'' particle, $\elh$. Provided the difference $[\Om]$ 
is small compared with the uniform crust angular velocity $\Om$ it can 
be seen from (\ref{fangm}) that these modified Taylor cylinders will differ
only by a small deformation from the ordinary Taylor cylinders that are 
characterised by fixed values of the cylindrical radial coordinate $\hh$.

It is interesting to consider what would happen if the force given by 
(\ref{forcec}) were also supposed to vanish,  so that there would be no 
net force at all: 
\be \ff^{\rmc}_{i}=0\Rightarrow A_\rmc\nabla_{\! i\,}{\cC}_\rmc
+\nabla_{\! i\,} \tilde\mmu^{\rmI}=0\, ,\fe
where  $A_\rmc$ is the nuclear mass number as introduced in
(\ref{nucmass}).  This would provide an analogue of the Proudman 
theorem, to the effect that the confined particle  energy parameter 
${\cC}_\rmc$ and the nuclear potential function $\tilde\mmu^{\rmI}$
 should both be uniform over surfaces characterised by a fixed value
of $A_\rmc$.

A noteworthy special case of the application of the foregoing theorems is 
to the scenario considered by Prix et al \cite{PrixComAnd02} in which there 
is no stratification, meaning that $\nabla_{\! i\,}\tilde\mmu^{\rmI}=0$ and 
for which the differential rotation is also postulated to be uniform so 
that one has $\nabla_{\! i\,}[\Om]=0$, in which case it follows that 
in the absence of any (pinning or external stress) force both 
${\cC}_\rmc$ and ${\cC}_\rmf$ would be uniform, as first integral 
constants of the motion, in the sense that their gradients too would 
vanish, $\nabla_{\! i\,}{\cC}_\rmc=\nabla_{\! i\,}{\cC}_\rmf=0$.

Assuming that (as well as the gravitational background $\phi$ that may 
have been provided by a Cowling type approximation) the distributions of 
the angular velocities (or the corresponding angular  momenta) and in our 
case also of the nuclear number density $\nI$ (or the corresponding 
stratification potential $\tilde\mmu^{\rmI}$) have been specified in 
advance to have values resulting from the previous history of the system, 
then the specification of two more independent quantities such as 
${\cC}_\rmc$ and ${\cC}_\rmf$ will suffice to determine the system 
completely, as remarked by Prix et al. \cite{PrixComAnd02}, who observed 
that their values as first integral constants could be fixed by the 
specification of the globally integrated totals of free and confined 
baryons.

A force free configuration with separate conservation of the free and
confined constituents is not however what is most realistic in a system
that is stationary over a secular timescale. The assumption of separate 
conservation of neutrons and protons is appropriate in the context of high
 frequency dynamical processes \cite{ComAnd01, AndComLan02} such as may
give rise to a two stream instability \citep {AndComPrix03}. However in
an effectively stationary context it is to be expected that transfusion 
(by beta processes) would be able to establish the chemical equilibrium 
condition (\ref{beteq}) to the effect that the affinity $\tilde{\cal A}$ 
should vanish, which by (\ref{endif}) means that one will have
\be {\cC}_\rmc={\cC}_\rmf+\elh\,[\Omega]\, .\label{nonuni}\fe

In the special case for which $[\Om]$ and hence (by the generalised 
Proudman theorem) also ${\cC}_\rmf$ are uniform it follows, since the 
distribution of $\elh$ will be highly non-uniform (roughly proportional 
to the squared cylindrical radius $\hh^2$) that  ${\cC}_\rmc$ will also 
be highly non-uniform. We thus obtain a scenario in which unlike that 
considered by Prix et al.\cite{PrixComAnd02} requires only a single 
constant of integration. This can be taken to be the globally integrated 
total baryon number, which suffices to determine the uniform value of 
${\cC}_\rmf$, from which ${\cC}_\rmc$ will be obtainable unambiguously 
(without any need to specify any other arbitrary parameter) as the 
non-uniform field provided by substituting the relevant fixed values of 
$[\Om]$ and ${\cC}_\rmf$. A scenario of this mathematically well defined 
type differs from the kind envisaged by  Prix et al. \cite{PrixComAnd02} 
in that, despite the fact that no vortex pinning is involved, it requires 
that the solid structure of the crust should provide a stress force density 
that according to (\ref{unpineq}) will be given as a sum of stratification 
and buoyancy deficit forces by
\be \ff_i=\nI\nabla_{\! i}\tilde \mmu^{\rmI}+
\nc[\Om]\nabla_{\!i}{\elh}\, .\label{freef}\fe
This could only vanish if $A_\rmc^{-1}\nabla_{\! i}\tilde\mmu^{\rmI}$
were equal in magnitude and opposite in sign to $[\Om]\nabla_{\!i}{\elh}$ 
which is mathematically conceivable but physically implausible (except near 
the stellar equator) because $\elh$ will depend mainly on the cylindrical 
coordinate $\hh$,  whereas one would expect that the other quantities 
involved would depend mainly on the spherical radial coordinate $\rr$. 

\subsection{Constants of integration for pinned superfluid case}

To deal with the case when the superfluid vortices are pinned to the 
``normal'' constituent it is convenient to focus attention not so much on
the comoving diagonal element ${\cC}_\rmf$ in the energy
matrix (\ref{ematrix}) but on the cross term
\be {\cC}={\calE}^\rmf_{\ \rmc}={\cC}_\rmf+[\vv^i]\,\muf_{\, i}
\, ,\label{crossed}\fe
which is interpretable as the energy per particle of the 
``free'' (superfluid) neutrons with respect to the``normal'' rest frame
of the ``confined'' baryons. Unlike the diagonal elements
$\cC_\rmf=-\pif_{\,\nu} \uu_\rmf^{\,\nu}$ and 
$\cC_\rmc=-\pic_{\,\nu}\uu_\rmc^{\,\nu}$ (and the other
off-diagonal element) this particular off-diagonal element
$\cC=-\pif_{\,\nu} \uu_\rmc^{\,\nu}$ has the property of 
being invariant with respect to the changes of chemical basis
that were discussed in Subsection \ref{chembas}.

The usual supposition that, instead of moving freely, as was supposed 
in the preceding subsection, the vortices will satisfy the pinning 
condition expressed by (\ref{vortrans}) and (\ref{pinning}) can be seen 
to be equivalent to the requirement that the ``free'' neutron current 
should be subject to a force density of the form
\be \ff^{\rmf}_{\,i}=-\ff^{\rm J}_{\,i} \fe
 where $\ff^{\rm J}_{\,i}$ is the Magnus force density reacting on the 
solid structure of the confined constituent, as  given
in accordance with the usual Joukowsky formula (see appendix) by
\be \ff^{\rm J}_{\,i}=\nf \varppi^{\rmf}_{ij}[\vv^j]\ ,\fe
which works out in this stationary circularly symmetric case as
\be \ff^{\rm J}_{\,i}=\nf[\Om]\nabla_{\! i}{\elh}\, .\label{Jouko}\fe
It is to be noticed that, unlike the buoyancy deficit force density
(\ref{boy}) to which it is proportional, this Magnus force density
is invariant with respect to changes of chemical basis. When 
substituted in the general formula (\ref{stress}) for the total force 
acting on the system it provides an expression of the simple form
\be \ff_{\,i}=\ff^{\chi}_{\,i}-\ff^{\rm J}_{\,i}\, .
\label{pineq}\fe

By substituting the Joukowsky formula in (\ref{forcef}) it can
be seen that in terms of the crossed particle energy (\ref{crossed}) 
the vortex pinning condition can be rewritten simply as
\be \nabla_{\! i\,} {\cC}= 0\, ,\fe
which means that for any pinned configuration $ {\cC}$ will be a 
uniform integral constant, whose value can be fixed by the total 
globally integrated baryon number. In terms of this constant 
${\cC}$ the comoving energy per particle of the ``free'' 
constituent will be obtainable from (\ref{crossed}) as the variable 
function of position given by
\be {\cC}_\rmf={\cC}-[\Om]\,\elh\, .\fe

It can be seen from (\ref{nonuni}) that the tranfusive equilibrium 
condition (\ref{beteq}) will be expressible simply as
\be {\cC}_\rmc={\cC}\, ,\fe
which means that,  when the specification of the system is completed
by the imposition of this chemical equilibrium condition, then
${\cC}_\rmc$ will also have a uniform constant value. 

The constancy of these quantities is interpretable as a modified 
Bernouilli theorem of a kind that is rather robust: it is to be 
emphasised that whereas the constancy deduced by Prix et al. 
\cite{PrixComAnd02} for ${\cC}_\rmc$ and ${\cC}_\rmc$ in the unpinned 
case was dependent on very restrictive assumptions about the uniformity 
of $[\Om]$ and the negligibility of stratification, no such restrictions 
are involved in the derivation of constancy for  ${\cC}_\rmf$ and 
${\cC}_\rmc$ in the pinned case. 

In this pinned equilibrium case, the analogue of the formula 
(\ref{freef}) for the stress force density that needs to be provided
by the solidity of the crust will be given, as a difference between 
stratification and Magnus-Joukowsky contributions by the expression
\be \ff_i=\nI\nabla_{\! i}\tilde \mmu^{\rmI}-
\nf[\Om]\nabla_{\!i}{\elh}\, ,\label{pinf}\fe
which, unlike  (\ref{freef}), has the noteworthy property of being 
chemically invariant. 

The work in this section has  used no approximations  apart from the 
postulates of exact stationarity and axisymmetry, but if we make the 
realistic supposition that the configurations under consideration differ
 only slightly from a rigidly rotating reference state with fixed angular 
velocity $\bar\Om$ then, to lowest order, we shall obtain the 
expressions 
\be \elh=\mm\,\bar\Om\, \hh^2\, ,\hskip 1 cm \nabla_{\!i}\elh=
2 \mm\,\bar\Om\, \hh\nabla_{\!i}\hh\, .\fe
This dependence of the angular momentum (per superfluid particle)
$\elh$ on the cylindrical radial coodinate $\hh$ contrasts with 
the dependence mainly on the spherical radial coordinate $\rr$ that one 
would expect for the number densities $\nI$ and $\nf$, as well as for the 
angular velocity difference $[\Om]$ (which, in view of (\ref{lag}), might 
be expected to be roughly proportional to the depth dependent ratio 
$\mm/\mm_\star$). This is why, as in the case of  (\ref{freef}), it is 
physically implausible that the gradient of the stratification potential  
$\tilde\mmu^{\rmI}$ could be adjusted so as to cancel the terms in 
(\ref{pinf}) except near the stellar equator.

\section{Conclusions}

The foregoing work shows how it is possible to construct an extensive 
range of simple, explicitly integrable, axisymmetric stratified two-fluid
neutron star models in which a small differential angular velocity 
$[\Om]$ can be chosen in advance as an arbitrary (implicitly history 
dependent) function of the spherical coordinates $\rr$ and $\ttheta$ in the 
pinned case, and as an arbitrary function just  of the cylindrical 
coordinate $\hh= \rr\, {\rm cos}\,\ttheta$ in the unpinned case to 
which the generalised Proudman theorem applies.

This idealised category of non-dissipative models includes, as extreme 
case, the kind of force-free model constructed by Prix et al. 
\cite{PrixComAnd02} on the basis of the postulate of separate conservation 
of protons and neutrons. Except in this special case it is shown that the 
equilibrium of the two-fluid model necessitates the presence of a balancing 
force (given by (\ref{freef}) for unpinned  configurations and by 
(\ref{pinf})  for pinned configurations) that must presumably be provided by 
stress in the solid structure of the crust (whose detailed treatment is 
beyond the scope of the purely fluid description used here) which can build 
up only to a certain point at which the breakdown responsible for the glitch 
phenomenon will occur.

The exceptional case in which the presence of such (glitch producing) 
stress was avoided \citep{PrixComAnd02} depended on the postulate of 
separate conservation of protons and neutrons, a condition that would be 
realistic in the short timescale context of dynamic perturbations but not 
in the long timescale context of stationary configurations, for which the 
postulate of chemical equilibrium is advocated here as a more plausible 
idealisation. 

It is to be understood that the force density given either by
(\ref{freef}) or more likely by (\ref{pinf}) will have to be 
balanced by a solid stress gradient, meaning that it will satisfy a 
relation of the form $\ff_i=\nabla_{\! j}\langle\SeS\rangle{^j_{\ i}}$ in 
which $\langle\SeS\rangle{^j_{\ i}}$ is a shear stress tensor that is the 
trace-free part of an elastic stress tensor $\SeS^j_{\ i}$ of the kind 
recently described  in the framework of an elastic solid  model that 
is already available~\citep{CCC05} for the description of the outer crust, 
of which an appropriate neutron conducting generalisation will soon be 
ready ~\citep{CC06} for application to the inner crust (with density above 
the neutron drip threshold) under consideration here. The idea is that 
this shear stress tensor $\langle\SeS\rangle_{ij}$ will be proportional 
to a trace-free shear strain tensor that can not excede a critical 
magnitude beyond which the solid structure will breakdown, and that 
a glitch will occur either when this critical value is reached or else, 
in the pinned case (\ref{pinf}), at what may be an earlier stage when 
the Joukowski-Magnus force density (\ref{Jouko}) reaches a vortex 
slippage limit at which unpinning occurs.

For the study of processes occurring on intermediate timescales it will 
of course be necessary to include allowance for the dissipative 
processes that have been neglected in the idealised models considered
here. As well as the controversial question of vortex slippage,
\citep{Jones98, JonesAnd01, DallOsso03,Donati03, Donati04} a particularly 
important issue concerns the timescales
\citep{Reisenegger97,Yakovlev01, Villain04} of the weak 
interaction processes leading to chemical equilibrium.

\section*{Acknowledgments}

The authors are grateful to Silvano Bonazzola, Loic Villain and Pawel Haensel
for discussions concerning stratification and deviations from chemical 
equilibrium in neutron stars. Nicolas Chamel acknowledges support from the 
Lavoisier program of the French Ministry of Foreign Affairs.

\section{Appendix}

\subsection{Joukowsky formula for Magnus force}
\label{Magnus}

The purpose of this appendix is to derive the formula used above 
for the force exerted on the fluid in the pinned case and to 
verify that it does indeed balance the collective effect of the 
Magnus forces exerted on the individual vortices. We shall start by 
evaluating the latter, using the generalised Joukowski formula that 
has recently been derived \citep{CC04} for the force exerted on an 
isolated vortex by a mixture of \emph{irrotational perfect} fluids.

The present application is to the case of a vortex that is fixed with 
respect to the crust rest frame characterised by the flow 4 vector 
$\uu_\rmc^{\,\mu}$ of the ``normal'' constituent, and that is oriented  
in the direction specified by a spacelike unit 4-vector $\nu^\mu$ say, 
for which the force per unit length will be given by
\be {\calF}_\nu =\calC^\rmf\nn^\mu\vvarepsilon_{\mu \nu\rho\sigma}
\nu^{\rho}\uu_\rmc^{\,\sigma}\fe
where $\nn^\mu$ is the (chemical gauge invariant) relative flow 
current vector of the superfluid
and $\calC$ is its (chemical gauge invariant) circulation as given by
\be {\calC}^\rmf = \oint \pif_{\,\mu} {\rm d} \xx^\mu\, , \label{A3}\fe 
in which the integral is carried out along any \emph{closed} path 
surrounding the vortex. 

Since the effect of many such vortices in closely spaced parallel array
will be represented by a vorticity vector 
\be \ww^\mu=\ww\, \nu^\mu\fe
of the type introduced in (\ref{wvector}), whose magnitude $\ww$
is the value of the total circulation per unit area, it follows
that the corresponding force density (obtained as force per unit length
per unit area) will be given by 
\be \ff^{\rm J}_{\, \nu}=\frac{\ww}{\calC^\rmf}\calF_\nu\fe
so we end up with the formula
\be \ff^{\rm J}_{\, \nu}=\nn^\mu\vvarepsilon_{\mu \nu\rho\sigma}
\ww^{\rho} \uu_\rmc^{\,\sigma}\label{Jou}\fe
in which it is to be observed that the value of the circulation
$\calC^\rmf$ round an individual vortex has cancelled out.
(Thus it is not necessary to know that it will actually
be given in terms of the Dirac-Planck constant by the formula
$\calC^\rmf=\pi\hhbar$ in which the usual factor $2$ is missing
because of the pairing due the fermionic nature of the neutrons.)

To obtain the force density on the fluid we start from the formula 
(\ref{3}) in which, since we are considering a stationary axisymmetric
 situation, the divergence term will automatically drop out leaving the 
expression
\be \ff^\rmf_{\,\mu} = \nf^\nu \varppi^{\rmf}{_{\!\nu\mu}}\, .
\label{A11}\fe
which would be equivalent just to the ordinary Euler equation if the
force term vanished.  It is evident that the pinning condition
\be u_{\rmc}^\nu\varppi^{\rmf}{_{\!\nu\mu}} = 0, \label{A12}\fe
can be used to rewrite  (\ref{A11}) in the form 
\be \ff^{\rmf}_{\,\mu}=\nf (\uu_\rmf^{\,\nu}-\uu_\rmc^{\,\nu})
\varppi^{\rmf}_{\nu\mu}=\nn^{\nu}\varppi^{\rmf}_{\,\nu\mu}
\, .\label{AA}\fe
It can also be seen to follow from (\ref{A12}) that, in terms of the 
vector introduced in (\ref{wvector}), the vorticity form will be 
expressible as
\be\varppi^{\rmf}_{\,\nu\mu}=
\vvarepsilon_{\mu \nu\rho\sigma}\ww^{\rho}\uu_\rmc^{\,\sigma}\, ,\fe
so that (\ref{Jou}) can be rewritten in the form
\be\ff^{\rm J}_{\,\mu}=-\nn^{\nu}\varppi^{\rmf}_{\,\nu\mu}
\, ,\fe
which evidently does indeed exactly balance (\ref{AA}).

\subsection{Effect of entrainment on vortex density}

It is of interest to consider the effect of entrainment on the 
vorticity magnitude $\ww$ which is the circulation per unit area
round an infinitesimal circuit orthogonal to the local axis of 
rotation. Remarking that the circulation $\calC^\rmf$ is simply 
given by $2\pi \elh$, where $\elh$ is the angular momentum per free 
particle, it can be seen from (\ref{fang}), by considering the 
circulation between nearby circuits with crylindrical radius values 
$\hh$ and $\hh+\delta\hh$, that the local vorticity magnitude will 
be expressible in terms of the angular velocity difference 
$[\Om]=\Of-\Oc$ and the mass increment $\mm^\rmf_{\,\rmc}
=\mm_\star-\mm$ as
\be \ww=\frac{\delta\calC^\rmf}{2\pi \hh \delta \hh}
= 2\mm\Of + 2 \mm^\rmf_{\,\rmc}   [\Om]+
\hh\frac{\delta(\mm_\star [\Om])}{\delta\hh}
 \, . \label{ww}\fe

Knowing that the circulation round an individual quantised vortex 
line in the fermionic superfluid condensate will be given by 
$\calC^\rmf=\pi\hhbar$ we see that the corresponding surface number 
density of quantised vortex lines will simply be given by 
$\ww/\calC^\rmf=\ww/(\pi\hhbar)$. The first term in equation 
(\ref{ww}) can be recognised as the well-known formula 
for a single superfluid.

In practice, the extra terms proportional to $[\Om]$ and its 
gradient will be negligible for the astrophysical applications under 
consideration, in which we shall generally have $[\Om]\ll\Om$, 
provided we are using an ``operational'' gauge in which magnitude of 
the effective mass  $\mm_\star$ will remain comparable with the 
ordinary baryonic mass $\mm$. If the evaluation were carried out in 
a ``comprehensive'' gauge, meaning a chemical basis in which all the 
neutrons are deemed to be ``free'', the relevant value of the effective 
mass would become extremely large near the ``neutron drip'' transition, 
but the value of the (gauge dependent) difference $[\Om]$ would 
correspondingly become even smaller, since the product $\mm_\star[\Om]$ 
is chemically invariant, so that the same value would ultimately be 
obtained for the vorticity magnitude $\ww$ itself.
 

\label{lastpage}

\end{document}